\documentclass[12pt]{iopart}
\eqnobysec

\usepackage{iopams,graphicx,amssymb}

\begin{document}

\title[Inertial-range behaviour of a passive scalar field in a random shear
flow]
{Inertial-range behaviour of a passive scalar field in a random shear flow:
Renormalization group analysis of a simple model}

\author{N.\,V. Antonov and A.\,V. Malyshev}

\address{Department of Theoretical Physics, St.~Petersburg State University
\\  St.~Petersburg--Petrodvorez, 198504 Russia}

\ead{nikolai.antonov@pobox.spbu.ru}

\begin{abstract}
Infrared asymptotic behaviour of a scalar field, passively advected by a
random shear flow, is studied by means of the field theoretic renormalization
group and the operator product expansion. The advecting velocity is Gaussian,
white in time, with correlation function of the form
$\propto \delta(t-t') / k_{\bot}^{d-1+\xi}$, where
$k_{\bot}=|{\bf k}_{\bot}|$ and  ${\bf k}_{\bot}$
is the component of the wave vector, perpendicular to the distinguished
direction (`direction of the flow') --- the $d$-dimensional generalization
of the ensemble introduced by Avellaneda and Majda [{\it Commun. Math. Phys.}
{\bf 131}: 381 (1990)]. The structure functions of the scalar field in the
infrared range exhibit scaling behaviour with exactly known critical
dimensions. It is strongly anisotropic in the sense that the dimensions
related to the directions parallel and perpendicular to the flow are
essentially different. In contrast to the isotropic Kraichnan's rapid-change
model, the structure functions show no anomalous (multi)scaling and have
finite limits when the integral turbulence scale tends to infinity. On the
contrary, the dependence of the internal scale (or diffusivity coefficient)
persists in the infrared range. Generalization to the velocity field with
a finite correlation time is also obtained. Depending on the relation
between the exponents in the energy spectrum
${\cal E} \propto  k_{\bot}^{1-\varepsilon}$
and in the dispersion law  $\omega \propto  k_{\bot}^{2-\eta}$,
the infrared behaviour of the model is given by the limits of vanishing
or infinite correlation time, with the crossover at the ray $\eta=0$,
$\varepsilon>0$ in the $\varepsilon$--$\eta$ plane.
The physical (Kolmogorov) point $\varepsilon=8/3$, $\eta=4/3$ lies inside
the domain of stability of the rapid-change regime;
there is no crossover line going through this point.

\end{abstract}


\maketitle

\section{Introduction} \label{sec:Intro}

The problem of turbulent advection, being of practical importance in itself,
has become a cornerstone in studying fully developed hydrodynamical
turbulence on the whole \cite{FGV}. On one hand, deviations from the
classical Kolmogorov theory --- intermittency and anomalous scaling
\cite{Legacy,Monin} --- are much stronger pronounced for a passively advected
scalar field (temperature of the fluid or concentration of impurity) than
for the advecting turbulent field itself. On the other, the problem of
passive advection appears easier tractable theoretically.
Most remarkable progress was achieved for Kraichnan's rapid-change model:
for the first time, the anomalous exponents were derived on the basis of a
dynamical model and within controlled approximations \cite{Falk1,GK}.

In Kraichnan's model, the turbulent velocity field is modelled by the
Gaussian distribution with the pair correlation function of the form
\begin{eqnarray}
\langle v_{i} v_{j} \rangle \propto D_{0}\, \delta(t-t')\,
P_{ij} \, k^{-d-\xi},
\label{OK}
\end{eqnarray}
where $P_{ij} = \delta _{ij} - k_i k_j / k^2$ is the transverse projector,
$k\equiv |{\bf k}|$ is the wave number, $D_{0}>0$ is an amplitude factor,
$d$ is the dimension of the ${\bf x}$ space and $\xi$ is an arbitrary
exponent. The latter can be viewed as a kind of H\"{o}lder's exponent,
which measures `roughness' of the velocity field; the `Batchelor limit'
$\xi\to2$ corresponds to smooth velocity, while the most realistic
(Kolmogorov) value is $\xi=4/3$ \cite{FGV}.

The issue of interest is, in particular, the behaviour of the equal-time
structure functions
\begin{equation}
S_{n}(r) = \langle\left[\, \theta(t,{\bf x})-\theta(t,{\bf x}')\,
\right]^{n} \rangle, \quad  r =|{\bf x}-{\bf x}'|
\label{struc}
\end{equation}
of the scalar field $\theta(t,{\bf x})$ in the inertial range
$\ell\ll r\ll{\cal L}$, where $\ell$ is the dissipation length and
${\cal L}$ is the integral turbulence scale.
Within the so-called zero-mode approach, developed in \cite{Falk1,GK},
it was shown that in the inertial range the functions (\ref{struc})
are independent of the diffusivity coefficient and have the forms:
\begin{equation}
S_{2n}(r) \propto D_{0}^{-n}  r^{n(2-\xi)}\, (r/{\cal L})^{\Delta_{n}},
\label{HZ1}
\end{equation}
with negative {\it anomalous exponents} $\Delta_{n}$, whose first terms of
the expansions in $1/d$ \cite{Falk1} and $\xi$ \cite{GK} are the following:
\begin{equation}
\Delta_{n} = -2n(n-1)\xi/d +O(1/d^{2}) = -2n(n-1)\xi/(d+2)+O(\xi^{2}).
\label{HZ3}
\end{equation}
Thus the functions (\ref{struc}) depend on the integral scale and diverge
for ${\cal L}\to\infty$, in contradiction with the classical Kolmogorov
theory.

In \cite{RG} and subsequent papers, the field theoretic renormalization
group (RG) and operator product expansion (OPE) were applied to Kraichnan's
model; see \cite{JphysA}  for the review and references. In the RG approach,
the exponent $\xi$ plays the part analogous to that played by
$\varepsilon=4-d$ in Wilson's theory of critical phenomena, while
$d$ remains a free parameter. The anomalous scaling for the structure
functions emerges as a consequence of the existence in the corresponding
operator product expansions of `dangerous' composite fields
(composite operators in the field theoretic terminology) of the form
$(\partial\theta)^{2n}$, whose {\it negative} critical dimensions are
identified with the anomalous exponents $\Delta_{n}$. This allows one
to construct a systematic perturbation expansion for the anomalous
exponents and to calculate them up to the orders $\xi^{2}$ \cite{RG}
and $\xi^{3}$ \cite{cube}.

In this paper, the RG+OPE approach is applied to the model of a passive
scalar field in a random shear flow: the Gaussian velocity field is oriented
along a fixed direction ${\bf n}$ (`direction of the flow') and depends
only on the coordinates in the subspace orthogonal to ${\bf n}$. In the
momentum space, its correlation function has the form simillar to (\ref{OK}):
$\langle vv \rangle \propto \delta(t-t')\,  k_{\bot}^{-d+1-\xi}$, where
$k_{\bot}= |{\bf k}_{\bot}|$ and ${\bf k}_{\bot}$ is the component of the
momentum ${\bf k}$ perpendicular to ${\bf n}$. This model can be viewed as
a $d$-dimensional generalization of the strongly anisotropic velocity
ensemble introduced in \cite{AM} in connection with the turbulent diffusion
problem and further studied and generalized
in a number of papers~\cite{AM1}--\cite{Walls}.

We show that the inertial-range behaviour of this model appears essentially
different from the isotropic Kraichnan's model: due to the absence of
dangerous composite operators, the structure functions (\ref{struc}) have
finite limits at ${\cal L}\to\infty$ and thus show no anomalous scaling in
the sense of (\ref{HZ1}). On the contrary, dependence on the diffusivity
(and thus on the dissipation length) persists in the inertial range.
Following the nomenclature of the monographs \cite{Legacy,Monin},
one can say that, in complete contradistinction with isotropic Kraichnan's
model, the first Kolmogorov hypothesis is valid in the present case, while
the second hypothesis is violated.

The paper is organized as follows. The sections
\ref{sec:QFT}--\ref{sec:OPE} are devoted to the rapid-change version
of the model (vanishing correlation time); generalization to the
finite-correlated case is given in section~\ref{sec:FCT}.

In section~\ref{sec:QFT} we give detailed description of the model, present
its field theoretic formulation and the corresponding diagrammatic technique.
In section~\ref{sec:Reno} we analyze canonical dimensions and ultraviolet
(UV) divergences of the model. We show that, after an appropriate extension,
the model becomes multiplicatively renormalizable. We derive the explicit
expression for the only independent renormalization constant, which is given
exactly by the one-loop approximation. In section~\ref{sec:FPS} we derive the
differential RG equations with exactly known coefficients ($\beta$ function
and anomalous dimensions $\gamma$) and show that they possess an infrared
(IR) attractive fixed point, which governs the scaling behaviour of the
Green functions in the IR range.

In section~\ref{sec:Dim} we present the corresponding critical dimensions
for the basic fields and parameters. Our model is {\it strongly} anisotropic
in the sense that, in contrast to previous RG+OPE studies of anisotropic
passive advection \cite{Uni}--\cite{Uni3}, it does not include parameters
that could be tuned to make the velocity statistics isotropic, and hence
it does not include the isotropic Kraichnan's model as a special case.
As an interesting consequence, the critical dimensions related to the
directions parallel and perpendicular to the flow are essentially different.

Section~\ref{sec:Operators} is devoted to the composite operators.
As already mentioned, the key role in the RG+OPE approach to anomalous
scaling is played by the dimensions of the Galilean invariant operators
$(\partial\theta)^{2n}$, built of the scalar gradients
\cite{RG}--\cite{cube}. In the isotropic case, there is only one such
operator for a given $n$, namely
$(\partial_{i}\theta\partial_{i}\theta)^{n}$. In the strongly anisotropic
case of a shear flow, there is a set of $(n+1)$ relevant operators for
each $n$. They mix heavily in renormalization and give rise to a set of
critical dimensions rather than a single $\Delta_{n}$. Nevertheless, it
turns out that exact expressions can be derived for these dimensions in
our model. Furthermore, in contrast to their counterparts (\ref{HZ3}) in
the isotropic Kraichnan's model, they all are positive.

Sections~\ref{sec:IRA} and~\ref{sec:OPE} apply the results of the preceding
analysis to the inertial-range asymptotic behaviour of the structure
functions (\ref{struc}). In section~\ref{sec:IRA} their behaviour in the
IR range $r \gg \ell$ is established; it turns out that those functions
retain the dependence on the UV scale $\ell$. The inertial range corresponds
to the additional condition that $r\ll{\cal L}$; it is studied by means of the
OPE in section~\ref{sec:OPE}. Due to the absence of relevant dangerous
operators with negative dimensions, the structure functions appear finite
for $(r/{\cal L})\to 0$ and thus show no anomalous scaling in the sense of
(\ref{HZ1}). The resulting inertial-range asymptotic expressions, presenting
the main outcome of this study, are summarized in (\ref{ScC})--(\ref{A2}).

In section~\ref{sec:FCT}, the generalization of the above results to the
velocity ensemble with finite correlation time is given. The energy spectrum
is taken in the form ${\cal E} \propto k_{\bot}^{1-\varepsilon}$,
while the dispersion law is $\omega \propto k_{\bot}^{2-\eta}$.
It is shown that the IR behaviour of the model is nearly exhausted by the
two limiting cases: the rapid-change type behaviour, realized for
$\varepsilon>\eta>0$ (with $\xi= \varepsilon - \eta>0$), and the frozen
(time-independent) behaviour, realized for $\varepsilon>0$, $\eta<0$. The
crossover line between the two regimes is the ray $\eta=0$, $\varepsilon>0$
in the $\varepsilon$--$\eta$ plane. In contrast to the isotropic case, where
the physical (Kolmogorov) point $\varepsilon=8/3$, $\eta=4/3$ lies exactly
on the crossover line between the rapid-change and frozen regimes
\cite{Chetak}--\cite{Juha2}, now this point lies deep inside the domain of
stability of the nontrivial rapid-change behaviour; there is no crossover
line going through this point. This result is in agreement with the findings
of the exact analysis of the $d=(1+1)$-dimensional case by~\cite{Glimm,Walls}
and in disagreement with \cite{AM}--\cite{AM2}; this issue is further
discussed in section~\ref{sec:Conc}, which is also reserved for conclusions.

\section{Description of the model and the field theoretic formulation}
\label{sec:QFT}

The advection-diffusion equation for the scalar field $\theta(x)$ with
$x=\{ t,{\bf x}\}$ has the form
\begin{eqnarray}
\nabla_{t} \theta = \nu_0 \partial^{2} \theta + \zeta,
\label{eq1}
\end{eqnarray}
where
\begin{eqnarray}
\nabla_{t} = \partial_{t} + v_{i} \partial_{i}
\label{nabla}
\end{eqnarray}
is the Galilean covariant (Lagrangian) derivative,
$\partial_{t} = \partial/ \partial t$,
$\partial_{i} = \partial/ \partial x_{i}$,
$\partial^{2} = \partial_{i}\partial_{i}$ is the Laplacian,
$\nu_{0}$ is the diffusion coefficient and $\zeta(t,{\bf x})$ is a
Gaussian random noise with zero mean and the pair correlation function
\begin{eqnarray}
\langle \zeta(t,{\bf x})\zeta(t',{\bf x}')  \rangle =
\delta(t-t')\, C({\bf r}), \quad  {\bf r}= {\bf x}-{\bf x}'.
\label{forceD}
\end{eqnarray}
The function $C({\bf r})$ is finite at ${\bf r}=0$ (and we assume the
normalization $C(0)=1$) and rapidly decays for ${\bf r}\to\infty$; its
precise form is inessential. For incompressible fluid, the velocity
field ${\bf v}= \{v_{i}(x)\}$ is transverse due to the continuity
relation: $\partial_{i} v_{i}=0$.

Let ${\bf n}$ be a unit constant vector that determines some distinguished
direction (`direction of the flow'). Then any vector can be decomposed
into the components perpendicular and parallel to the flow, for example,
${\bf x} = {\bf x}_{\bot} + {\bf n}\, x_{\parallel}$ with
${\bf x}_{\bot} \cdot {\bf n} =0$.
The velocity field will be taken in the form
\begin{eqnarray}
{\bf v} = {\bf n} v(t, {\bf x}_{\bot}),
\label{vello}
\end{eqnarray}
where $v(t, {\bf x}_{\bot})$ is a scalar function independent of
$x_{\parallel}$.
Then the incompressibility condition is automatically satisfied:
\begin{eqnarray}
\partial_{i} v_{i} = \partial_{\parallel} v(t, {\bf x}_{\bot}) = 0.
\label{inko}
\end{eqnarray}
For $v(t, {\bf x}_{\bot})$ we assume a Gaussian distribution with zero
mean and the pair correlation function of the form:
\begin{eqnarray}
\langle v(t,{\bf x}_{\bot}) v(t', {\bf x}_{\bot}') \rangle =
\delta(t-t') \int \frac{d {\bf k}}{(2\pi)^{d}} \,
\exp \left\{ {\rm i} {\bf k}\cdot ({\bf x}-{\bf x}') \right\} D_{v} (k)=
\nonumber \\
= \delta(t-t') \int \frac{d {\bf k}_{\bot}}{(2\pi)^{d-1}} \, \exp
\left\{ {\rm i} {\bf k}_{\bot}\cdot ({\bf x}_{\bot}-{\bf x}'_{\bot})
\right\} \widetilde D_{v} (k_{\bot}) , \quad  k_{\bot}=|{\bf k}_{\bot}|
\label{veloc1}
\end{eqnarray}
with the scalar coefficient functions 
\begin{eqnarray}
D_{v} (k)= 2\pi \delta(k_{\parallel}) \, \widetilde D_{v} (k_{\bot}) ,
\quad \widetilde D_{v} (k_{\bot}) = D_{0}\, k_{\bot}^{-d+1-\xi}.
\label{veloc2}
\end{eqnarray}
Here and below $d$ is the dimension of the ${\bf x}$ space, $D_{0}>0$
is a constant amplitude factor and $\xi$ an arbitrary exponent.
The IR regularization
in (\ref{veloc1}) is provided by the cutoff $k_{\bot}>m$, where
$m \sim {\cal L}^{-1}$ is the reciprocal of the integral turbulence scale.
Its precise form is inessential; the sharp cutoff is the most convenient
choice from the calculational viewpoints. The natural interval for the
exponent is $0< \xi <2 $, when the so-called `effective eddy diffusivity'
\begin{eqnarray}
{\cal V}({\bf r}_{\bot}) = \int \frac{d {\bf k}_{\bot}}{(2\pi)^{d-1}} \,
\left\{ 1- \exp \left( {\rm i} {\bf k}_{\bot}\cdot {\bf r}_{\bot} \right)
\right\} \, \widetilde D_{v} (k_{\bot})
\label{effect}
\end{eqnarray}
has a finite limit for $m\to0$; it includes the most realistic Kolmogorov
value $\xi=4/3$.

In order to ensure multiplicative renormalizability of the model, it is
necessary to split the Laplacian in (\ref{eq1}) into the parallel and
perpendicular parts $\partial^{2} \to \partial^{2}_{\bot}
+ f_{0} \partial^{2}_{\parallel}$ by introducing a new parameter
$f_{0}>0$. Here $\partial^{2}_{\bot}$ is the Laplacian in the subspace
orthogonal to the vector ${\bf n}$ and
$ \partial_{\parallel} = \partial/ \partial x_{\parallel}$.
In the anisotropic case, these two terms will be renormalized
in a different way. Thus equation (\ref{eq1}) becomes
\begin{eqnarray}
\nabla_{t} \theta =  \nu_0\, \left\{  \partial^{2}_{\bot} + f_{0}
\partial^{2}_{\parallel} \right\} \theta + \zeta;
\label{eq2}
\end{eqnarray}
this completes formulation of the model.
It remains to note that, for the velocity field (\ref{vello}),
the covariant derivative in (\ref{nabla}) takes on the form
\begin{eqnarray}
\nabla_{t} = \partial_{t} + v(t,{\bf x}_{\bot}) \partial_{\parallel}.
\label{nabla2}
\end{eqnarray}

Interpretation of the splitting of the Laplacian term in (\ref{eq2}) can
be twofold. On one hand, stochastic models of the type (\ref{eq1}) are
phenomenological and, by construction, they must include all
the IR relevant terms allowed by symmetry. The fact that the splitting is
required by the renormalization procedure means that it is not forbidden by
dimensionality or symmetry considerations and, therefore, it is natural to
include the general value $f_{0}\ne1$ to the model from the very beginning.
On the other hand, one can insist on studying the original model with
$f_{0} =1$ and $O_{d}$ covariant Laplacian term, although that symmetry is
broken to $O_{d-1}\otimes Z_{2}$ by the interaction with the anisotropic
velocity ensemble ($Z_{2}$ is the reflection symmetry
$x_{\parallel} \to - x_{\parallel}$).
Then the extension of the model to the case $f_{0}\ne1$ can be
viewed as a purely technical trick which is only needed to ensure the
multiplicative renormalizability and to derive the RG equations. The latter
should then be solved with the special initial data corresponding to
$f_{0} =1$ (in renormalized variables this anyway will correspond to
general initial data with $f\ne 1$). Since the IR attractive fixed point
of the RG equations is unique (see section~\ref{sec:FPS}), the resulting
IR behaviour will be the same as for the general case of the extended
model with $f_{0}\ne1$.

According to the general theorem (see e.g. chap.~5 of the monograph
\cite{Book3}), our stochastic problem is equivalent to the field theoretic
model of the extended set of fields
$\Phi = \{ \theta', \theta, {\bf v} \}$ with action functional
\begin{eqnarray}
{\cal S} (\Phi) = \frac{1}{2} \theta' D_{\zeta}\theta' +
\theta' \left\{ - \nabla_{t}  +  \nu_0\,
\left(  \partial^{2}_{\bot} + f_{0}
\partial^{2}_{\parallel} \right) \right\} \theta
+ {\cal S}_{v} ({\bf v}),
\label{action}
\end{eqnarray}
where $D_{\zeta}$ is the correlator (\ref{forceD}). The first few terms
represent the De Dominicis--Janssen action functional for the stochastic
problem (\ref{eq1}), (\ref{forceD}) at fixed ${\bf v}$; it involves
auxiliary scalar response field $\theta'(x)$.  All the required integrations
over $x=\{t,{\bf x}\}$ are implied, for example, the coupling term in
(\ref{action}), stemming from the derivative (\ref{nabla2}), in the
detailed notation has the form:
\begin{eqnarray}
-\theta' (v \partial_{\parallel}) \theta = - \int dt \int d x_{\parallel}
\int d {\bf x}_{\bot}
\theta'(x)\, v(t,{\bf x}_{\bot})\, \partial_{\parallel} \theta (x).
\label{Vi}
\end{eqnarray}
Due to the independence of the velocity field on the longitudinal
coordinate ${x}_{\parallel}$, the derivative in (\ref{Vi}) can
also be moved onto the field $\theta'$ using integration by parts:
\begin{eqnarray}
-\theta' (v \partial_{\parallel}) \theta =  \theta
(v \partial_{\parallel}) \theta '.
\label{Villon}
\end{eqnarray}
The last term in (\ref{action}) corresponds to the Gaussian averaging over
${\bf v}$ with correlator (\ref{veloc1}) and has the form
\begin{eqnarray}
{\cal S}_{v} ( {\bf v}) = - \frac{1}{2}\,
\int dt \int d{\bf x}_{\bot} d{\bf x}_{\bot}' v(t,{\bf x}_{\bot})
\widetilde D^{-1}_{v} ({\bf x}_{\bot}-{\bf x}'_{\bot}) v(t,{\bf x}_{\bot}'),
\label{Sv}
\end{eqnarray}
where
\begin{eqnarray}
\widetilde D^{-1}_{v} ({\bf r}_{\bot}) \propto D_{0}^{-1}
\, r_{\bot}^{2(1-d)-\xi}
\label{Dv}
\end{eqnarray}
is the kernel of the inverse linear operation $D^{-1}_{v}$ for the
correlation function $D_{v}$ in (\ref{veloc2}).

This formulation means that statistical averages of random quantities
in the original stochastic problem coincide with the Green functions of the
field theoretic model with action (\ref{action}), given by functional
averages with the weight $\exp {\cal S}(\Phi)$. This allows one to apply
the field theoretic renormalization theory and renormalization group to
our stochastic problem.

The action (\ref{action}) corresponds to the Feynman diagrammatic technique
with three bare propagators: the correlator of the velocity field
$\langle vv \rangle_{0}$, given by (\ref{veloc1}), (\ref{veloc2}), the
scalar Green function (in the frequency--momentum and time--momentum
representations):
\begin{eqnarray}
\langle \theta \theta' \rangle_{0} =
\left\{-{\rm i} \omega+ \epsilon({\bf k}) \right\}^{-1}
\ \leftrightarrow\ \Theta(t-t')\,
\exp \{- \epsilon({\bf k}) (t-t') \}
\label{lines3}
\end{eqnarray}
and the correlator of the scalar field
\begin{eqnarray}
\langle \theta \theta \rangle_{0} =
\frac{C({\bf k})}{\omega^{2}+ \epsilon^{2}({\bf k})}
\ \leftrightarrow\
\frac { C({\bf k})} {2\epsilon({\bf k})} \,
\exp \left\{- \epsilon({\bf k}) |t-t'| \right\} .
\label{lines2}
\end{eqnarray}
Here $C({\bf k})$ is the Fourier transform of the function from
(\ref{forceD}),
$\epsilon({\bf k}) = \nu_0 \big( k_{\bot}^{2}+f_{0} k_{\parallel}^{2}
\big)$ and $\Theta(\dots)$ is the Heaviside step function, so that the
function (\ref{lines3}) is retarded. The only vertex (\ref{Villon})
corresponds to the vertex factor
\begin{eqnarray}
V({\bf p}) = - {\rm i} p_{\parallel} = {\rm i} k_{\parallel} \, ,
\label{vertex}
\end{eqnarray}
where ${\bf p}$ is the momentum argument of the field $\theta$ and ${\bf k}$
is the momentum of $\theta'$.
The role of the bare coupling constant (expansion parameter in the ordinary
perturbation theory) is played by the parameter $w_{0}$, defined by the
relation
\begin{eqnarray}
D_{0}= w_{0} \nu_{0} f_{0}, \quad w_{0} \sim \Lambda^{\xi}
\label{g0}
\end{eqnarray}
with $D_{0}$ from (\ref{veloc2}). The last relation, following from
dimensionality considerations, sets in the typical UV momentum scale
$\Lambda \sim 1/\ell$, the reciprocal of the UV length scale.

\begin{figure}
\begin{center}
\includegraphics[width=13cm]{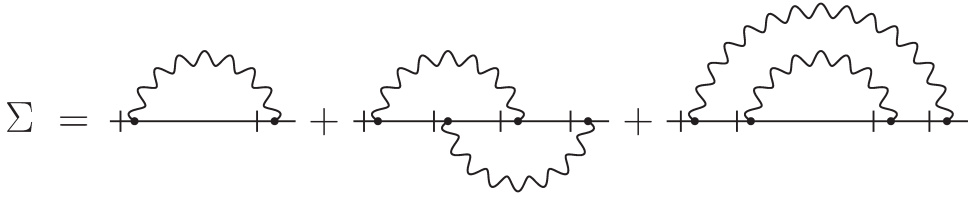}
\caption{\label{fig:sigma}
The self-energy operator (\protect\ref{Dy1}) in the two-loop approximation.}
\end{center}
\end{figure}

As an example, on figure~\ref{fig:sigma} we show the two-loop approximation
of the self-energy operator $\Sigma$ which enters the Dyson equation
\begin{equation}
\langle\theta'\theta\rangle_{\rm 1-ir} (\omega, {\bf p}) =
- \left\{ -{\rm i}\omega + \nu_{0} p_{\bot}^{2} +
\nu_{0} f_{0}  p^{2}_{\parallel} \right\} + \Sigma(\omega,{\bf p})
\label{Dy1}
\end{equation}
for the 1-irreducible Green function $\langle\theta'\theta\rangle_{\rm 1-ir}$
in the frequency--momentum representation.
The wavy lines denote the velocity correlator $\langle vv \rangle_{0}$,
while the solid lines correspond to the function
$\langle \theta \theta' \rangle_{0}$; the slashes mark the field $\theta'$.

\section{Canonical dimensions, UV divergences and renormalization}
\label{sec:Reno}

The analysis of UV divergences is based on the analysis of canonical
dimensions of the 1-irreducible Green functions. In general,
dynamic models have two scales: canonical dimension of some
quantity $F$ (a field or a parameter in the action functional) is completely
characterized by two numbers, the frequency dimension $d_{F}^{\omega}$
and the momentum dimension $d_{F}^{k}$; see e.g. chap.~5 in \cite{Book3}.
They are determined such that
$[F] \sim [T]^{-d_{F}^{\omega}} [L]^{-d_{F}^{k}}$, where $L$ is some
length scale and $T$ is the time scale.

Our strongly anisotropic model, however, has two independent length
scales, related to the directions perpendicular and parallel to the
vector ${\bf n}$, and requires a more detailed specification of the
canonical dimensions. Namely, one has to introduce two independent momentum
canonical dimensions $d_{F}^{\bot}$ and $d_{F}^{\parallel}$ so that
\[ [F] \sim [T]^{-d_{F}^{\omega}}  [L_{\bot}]^{-d_{F}^{\bot}}
[L_{\parallel}]^{-d_{F}^{\parallel}}, \]
where $L_{\bot}$ and $L_{\parallel}$ are (independent) length scales in the
corresponding subspaces. The dimensions are found from the obvious
normalization conditions $d_{k_{\bot}}^{\bot}= -d_{\bf x_{\bot}}^{\bot}=1$,
$d_{k_{\bot}}^{\parallel}=-d_{\bf x_{\bot}}^{\parallel}=0$,
$d_{k_{\bot}}^{\omega} = d_{k_{\parallel}}^{\omega}=0$,
$d_{\omega }^{\omega }=-d_t^{\omega }=1$, and so on, and from the
requirement that each term of the action functional (\ref{action})
be dimensionless (with respect to all the three independent dimensions
separately). The total momentum dimension can be found from the
relation $d_{F}^{k} = d_{F}^{\bot}+ d_{F}^{\parallel}$.
Then, based on $d_{F}^{k}$ and $d_{F}^{\omega}$, one can introduce the
total canonical dimension $d_{F}=d_{F}^{k}+2d_{F}^{\omega}  =
d_{F}^{\bot} + d_{F}^{\parallel} +2d_{F}^{\omega}$ (in the free theory,
$\partial_{t}\propto\partial^{2}_{\bot} \propto \partial^{2}_{\parallel}$),
which plays in the theory of renormalization of dynamic models the same
role as the conventional (momentum) dimension does in static problems.

\begin{table}
\caption{Canonical dimensions of the fields and parameters
in the model (\protect\ref{action})}
\label{table1}
\begin{tabular}{ccccccccc}
\br
$F$ & $\theta' $ & $\theta$ & $ {\bf v} $ &  $m,\mu, \Lambda $ &
$\nu ,\nu_{0}$ & $f, f_{0}$ & $w_{0}$ & $w$ \\
\br
$d_{F}^{\omega}$ & 1/2 & $-1/2$ & 1 & 0 & 1 & 0 & 0&  0\\
\mr
$d_{F}^{\parallel}$ & 1 & 0 & $-1$ & 0 & 0 & $-2$ & 0 & 0 \\
\mr
$d_{F}^{\bot}$ & $d-1$ & 0 & 0 & 1 & $-2$ & 2 & $\xi$ & 0 \\
\mr
$d_{F}^{k} = d_{F}^{\parallel} + d_{F}^{\bot}$ & $d$ & 0 & $-1$ & 1 &
$-2$ & 0 & $\xi$ & 0 \\
\mr
$d_{F} = d_{F}^{k} + 2d_{F}^{\omega}$ & $d+1$ & $-1$ & 1 & 1 & 0 & 0 &
$\xi$ & 0 \\
\br
\end{tabular}
\end{table}

The canonical dimensions of the model (\ref{action}) are given in
table~\ref{table1}, including renormalized parameters, which will be
introduced a bit later. From table~\ref{table1} it follows that our model
is logarithmic (the coupling constant $w_{0} \sim [L_{\bot}]^{-\xi}$
is dimensionless) at $\xi=0$, so that the UV divergences manifest
themselves as poles in $\xi$ in the Green functions.

The total canonical dimension of an arbitrary 1-irreducible Green function
$\Gamma = \langle\Phi \dots \Phi \rangle _{\rm 1-ir}$ is given by the
relation
\begin{equation}
d_{\Gamma }= d+2- \sum_{\Phi} N_{\Phi }d_{\Phi} = d+2-
N_{\theta'} d_{\theta'} - N_{\theta} d_{\theta} - N_{v} d_{v}.
\label{dGamma}
\end{equation}
Here $N_{\Phi}=\{N_{\theta},\,N_{\theta'},\,N_{v}\}$ are the numbers of
corresponding fields entering the function $\Gamma$, and the summation
over all types of the fields in (\ref{dGamma}) and analogous formulas below
is always implied.

Superficial UV divergences, whose removal requires counterterms, can be
present only in those functions $\Gamma$ for which the `formal index of
divergence' $d_{\Gamma}$ is a nonnegative integer.
Dimensional analysis should be augmented by the following observations:

(1) In any dynamical model of the type (\ref{action}), 1-irreducible
diagrams with $N_{\theta'}=0$ contain closed contours of retarded
propagators (\ref{lines3}) and therefore vanish.

(2) For any 1-irreducible Green function $N_{\theta'}- N_{\theta}=2N_{0}$,
where $N_{0}\ge0$ is the total number of the bare propagators
$\langle \theta \theta \rangle _0$ entering into any of its diagrams. This
fact is easily checked for any given function; it is illustrated by the
function (\ref{Dy1}) with $N_{\theta'}= N_{\theta}=1$ and $N_{0}=0$; see
figure~\ref{fig:sigma}. Obviously, no diagrams with $N_{0}<0$ can
be constructed. Therefore, the difference $N_{\theta'}- N_{\theta}$ is an
even nonnegative integer for any nonvanishing function.

(3) The derivative $\partial_{\parallel}$ at the vertex
$\theta' v \partial_{\parallel}\theta$ can be moved onto the field $\theta'$
due to the transversality of $v$, see (\ref{Villon}). Therefore, in any
1-irreducible diagram it is always possible to move the derivative onto any
of the external `tails' $\theta$ or $\theta'$, which reduces the real
index of divergence: $d_{\Gamma}' = d_{\Gamma}- N_{\theta}-N_{\theta'}$.
The fields $\theta$, $\theta'$ enter into the counterterms only in the form
of derivatives $\partial_{\parallel}\theta$, $\partial_{\parallel}\theta'$.

From table~\ref{table1} and (\ref{dGamma}) we find
\begin{equation}
d_{\Gamma}= d+2 - (d+1) N_{\theta'} + N_{\theta} -N_{v}, \quad
d_{\Gamma}\!\!' =(d+2)(1-N_{\theta'}) - N_{v}.
\label{IndeX}
\end{equation}
From these expressions we conclude that for any $d$, superficial
divergences can be present only in the 1-irreducible functions
$\langle\theta'\theta\dots\theta\rangle_{\rm 1-ir}$ with $N_{\theta'}=1$
and arbitrary
$N_{\theta}$, for which $d_{\Gamma}=2$, $d_{\Gamma}'=0$. However, all
functions with $N_{\theta}> N_{\theta'}$ vanish (see above) and obviously
do not require counterterms. We are left with the only superficially
divergent function $\langle\theta'\theta\rangle_{\rm 1-ir}$; the
corresponding
counterterm must contain two symbols $\partial_{\parallel}$ and therefore
reduces to $\theta' \partial_{\parallel}^{2}  \theta$.

Inclusion of this counterterm is reproduced by the multiplicative
renormalization of the action (\ref{action}) with the only
independent renormalization constant $Z_{f}$:
\begin{equation}
\nu_0=\nu, \quad f_0= f Z_{f}, \quad w_{0}= w\mu^{\xi }Z_{w},
\quad Z_{w}=Z_{f}^{-1}.
\label{18}
\end{equation}
Here the reference scale $\mu$ is an additional parameter of the
renormalized theory, $\nu$, $f$ and $w$ are renormalized analogs of
the bare parameters (with the subscript `0') and $Z=Z(w,\xi,d)$ are the
renormalization constants. Their relation in (\ref{18}) results from the
absence of renormalization of the contribution with $D_{0}$ in
(\ref{action}), so that
\begin{equation}
D_{0} = w_{0}\nu_0 f_{0} = w \mu^{\xi}\nu f,
\label{RenD}
\end{equation}
see (\ref{g0}).
No renormalization of the fields and the parameter $m$ is required:
\begin{equation}
m_{0}=m, \quad Z_{m}=1, \quad Z_{\Phi}=1 \quad {\rm for\ all}\ \Phi.
\label{NoRen}
\end{equation}
Here and below we use
the minimal subtraction (MS) scheme, where all renormalization constants
have the forms `1 + only poles in $\xi$.'

The constant $Z_{f}$ is determined by the requirement that the function
$\langle\theta'\theta\rangle_{\rm 1-ir}$, expressed in renormalized
variables, be UV finite, that is, finite at $\xi=0$. We recall that the
correlator $\langle vv \rangle_{0}$ contains the $\delta$ function
in time, while the propagator (\ref{lines3}) contains the step function.
Thus all the multiloop diagrams in the self-energy operator $\Sigma$ in
(\ref{Dy1}) contain self-contracted chains of the step functions,
like e.g. $\Theta(t_{1}-t_{2})\Theta(t_{2}-t_{3})\Theta(t_{3}-t_{1})$,
and therefore vanish. (In the frequency representation, all the integrands
have the poles in $\omega$ only in the lower complex half-plane.) This means
that the functions $\Sigma$ and $\langle\theta'\theta\rangle_{\rm 1-ir}$
are given {\it exactly} by the one-loop approximation.

The analytic expression for the only one-loop diagram has the form
\begin{eqnarray}
\Sigma({\bf p})= \int \frac{d\omega}{2\pi} \int \frac{d{\bf k}}{(2\pi)^{d}}
\, D_{v} (k) \, \frac{ {\rm i} p_{\parallel} {\rm i} (p-k)_{\parallel}}
{  -{\rm i} \omega + \epsilon({\bf p}-{\bf k})}
\label{D2}
\end{eqnarray}
with $D_{v}$ from (\ref{veloc2}) and $\epsilon({\bf k})$ from (\ref{lines2}).
This expression is independent of the external frequency. The prefactor,
coming from the vertices (\ref{vertex}), can be replaced with
$-p_{\parallel}^{2}$ due to the presence of the factor
$\delta(k_{\parallel})$ in (\ref{veloc2}); this is the diagrammatic analog
of the relation (\ref{Villon}). Integration over $\omega$ involves the
indeterminacy
\begin{eqnarray}
\int \frac{d\omega}{2\pi} \, \frac{1}{-{\rm i} \omega +
\epsilon ({\bf p}-{\bf k})} =  \Theta(0),
\label{inde}
\end{eqnarray}
the step function at the origin; it should be carefully resolved. In our
case, the function $\delta(t-t')$ should be understood as the limit of a
narrow function which is necessarily symmetric in $t \leftrightarrow t'$,
because (\ref{veloc1}) is a pair correlation function. Thus the quantity
in (\ref{inde}) must be unambiguously defined by half the sum of the limits:
$\Theta(0)=1/2$.

The integration over $k_{\parallel}$ is trivial due to the presence of the
factor $\delta(k_{\parallel})$ in (\ref{veloc2}). The result has the form
\begin{eqnarray}
\Sigma({\bf p})= - p\,{}_{\parallel}^{2} \, \frac{D_{0}}{2} \,
\int_{k_{\bot}>m} \frac{d {\bf k}_{\bot}}{(2\pi)^{(d-1)}} \,
\frac{1}{k_{\bot}^{d-1+\xi}}.
\label{D2a}
\end{eqnarray}
The remaining integration over ${\bf k}_{\bot}$ gives
\begin{eqnarray}
\Sigma({\bf p})= - p\,{}_{\parallel}^{2} \, D_{0} m^{-\xi}
\, \frac{S_{d-1}}{2(2\pi)^{(d-1)}} \frac{1}{\xi} ,
\label{D2b}
\end{eqnarray}
where $S_d=2\pi^{d/2}/\Gamma(d/2)$ with Euler's Gamma function is the
surface area of the unit sphere in the $d$-dimensional space.
Substituting the expression (\ref{D2b}) into the Dyson equation (\ref{Dy1})
and passing to renormalized parameters with the aid of relations (\ref{18})
and (\ref{RenD}) gives
\begin{equation}
\langle\theta'\theta\rangle_{\rm 1-ir} (\omega, {\bf p})
=  -\left\{ -{\rm i}\omega + \nu p_{\bot}^{2} +
\nu f Z_{f} p^{2}_{\parallel} \right\}
- p\,{}_{\parallel}^{2} \frac{wf\nu S_{d-1}}{2(2\pi)^{(d-1)}}
\left( \frac{\mu}{m} \right) ^{\xi} \frac{1}{\xi}.
\label{Dy2}
\end{equation}
In order to cancel the pole in $\xi$, the renormalization constant in the
MS scheme has to be chosen in the form
\begin{equation}
Z_{f} = 1 - \frac{w}{\xi},
\label{Zf}
\end{equation}
where we have absorbed the factor $S_{d-1}/2(2\pi)^{(d-1)}$ into
the coupling constant.

\section{RG equations and the fixed point} \label{sec:FPS}

Let $G(e_{0},\dots)$ be some correlation function in the original model
(\ref{action}) and $G_{R}(e,\mu,\dots)$ its analog in the renormalized
theory.
Here $e_{0}$ is the complete set of bare parameters, $e$ is the set
of their renormalized counterparts, and the ellipsis stands for the
other variables like the coordinates/momenta and times/frequences.
These functions differ only by
normalization and the choice of variables,  $G(e_{0})= Z_{G} G_{R}(e,\mu)$,
and can equally be used for the analysis of critical behaviour. (For the
correlation functions of the primary fields $Z_{G}=1$ due to the absence
of their renormalization, but it is instructive to discuss a more general
case.) We use
$\widetilde{\cal D}_{\mu}$ to denote the differential operation
$\mu\partial_{\mu}$ for fixed $e_{0}$ and operate on both sides of the
last equality with it. This gives the basic RG equation:
\begin{equation}
\left\{ {\cal D}_{RG} + \gamma_{G} \right\} \,G^{R}(e,\mu,\dots) = 0,
\label{RG1}
\end{equation}
where ${\cal D}_{RG}$ is the operation $\widetilde{\cal D}_{\mu}$
expressed in the renormalized variables:
\begin{equation}
{\cal D}_{RG}= {\cal D}_{\mu} + \beta\partial_{w} - \gamma_{f}{\cal D}_{f}.
\label{RG2}
\end{equation}
Here we have written ${\cal D}_{x}\equiv x\partial_{x}$ for any variable
$x$, and the RG functions (the $\beta$ function and the anomalous
dimensions $\gamma$) are defined as
\begin{equation}
\beta(w) = \widetilde {\cal D}_{\mu} w = w\,[-\xi-\gamma_{w}],
\label{RGF}
\end{equation}
\begin{equation}
\gamma_{e}(u) = \widetilde {\cal D}_{\mu} \ln Z_{e} = \beta \partial_{u}
\ln Z_{e} \quad {\rm for\ any\ quantity} \ e.
\label{RGF1}
\end{equation}
From the relations (\ref{18})--(\ref{NoRen}) it follows
\begin{equation}
\gamma_{\nu}= \gamma_{m}= \gamma_{\Phi}=0 \quad {\rm and} \quad
\gamma_{f}=-\gamma_{w},
\label{Nu}
\end{equation}
while from (\ref{RGF}) and (\ref{RGF1}) for $e=w$ one easily derives
\begin{equation}
\gamma_{w} (u) = \frac{- \xi {\cal D}_{w} \ln Z_{w}}
{1+{\cal D}_{w} \ln Z_{w}}, \quad
\beta (w)= \frac{- \xi w} {1+{\cal D}_{w} \ln Z_{w}} .
\label{Beta}
\end{equation}
Substituting (\ref{Zf}) into (\ref{Beta}) one obtains exact expressions
\begin{equation}
\gamma_{f} = - \gamma_{w} = w, \quad \beta=w[-\xi+w].
\label{BetaE}
\end{equation}
It is well known that IR asymptotic behaviour of the Green functions is
governed by IR attractive fixed points of the RG equations, defined by the
relations $\beta(w_{*})=0$ and $\beta'(w_{*})>0$. From (\ref{BetaE})
it follows that our model has a fixed point
\begin{equation}
w_{*} = \xi, \quad \beta'(w_{*}) = \xi
\label{FP}
\end{equation}
which is positive and IR attractive for $\xi>0$. At this point
\begin{equation}
\gamma_{f}^{*} = - \gamma_{w}^{*} = \xi.
\label{Anom}
\end{equation}
Here and below we denote $\gamma_{e}^{*}=\gamma_{e} (w_{*})$. We also stress
that all the expressions in (\ref{FP}) and (\ref{Anom}) are exact.

\section{Critical scaling and critical dimensions} \label{sec:Dim}

In the leading order of the IR asymptotic behaviour the Green functions
satisfy the RG equation with the substitution $w\to w_{*}$, which gives
\begin{equation}
\left\{ {\cal D}_{\mu} - \gamma_{f}^{*}{\cal D}_{f}
+ \gamma_{G}^{*} \right\} \,G^{R}(e,\mu,\dots) = 0.
\label{RGFP}
\end{equation}
Canonical scale invariance of the function $G^{R}$ with respect to
the three independent canonical dimensions (see section~\ref{sec:Reno})
can be expressed by the differential equations of the form
\begin{equation}
\left\{  \sum_{F} d^{*}_{F}\, {\cal D}_{F}
- d^{*}_{G} \right\} \, G^{R}(e,\mu,\dots) = 0,
\label{Canon}
\end{equation}
where the sum runs over all arguments of the Green function, including
the coordinates/momenta and times/frequences, and
$d^{*}_{F} = d_{F}^{\omega}$, $d_{F}^{\bot}$ and $d_{F}^{\parallel}$
are the corresponding canonical dimensions. In the time--coordinate
representation
\begin{equation}
d^{*}_{G}= \sum_{\Phi} N_{\Phi}d^{*}_{\Phi} = N_{\theta'}d^{*}_{\theta'}+
N_{\theta}d^{*}_{\theta}+ N_{v}d^{*}_{v},
\label{Ca}
\end{equation}
where $N_{\Phi}$ are the numbers of the fields entering the Green function,
cf. equation~(\ref{dGamma}). From table~\ref{table1} we find
\begin{eqnarray}
\left\{ -{\cal D}_{t} + {\cal D}_{\nu} - d^{\omega}_{G} \right\} \,
G^{R}(e,\mu,\dots) &=& 0,
\nonumber \\
\left\{ -{\cal D}_{\bot} + {\cal D}_{\mu} -2{\cal D}_{\nu}
+ 2{\cal D}_{f} + {\cal D}_{m} - d^{\bot}_{G} \right\} \,
G^{R}(e,\mu,\dots) &=& 0,
\nonumber \\
\left\{ -{\cal D}_{\parallel} - 2{\cal D}_{f} - d^{\parallel}_{G}
\right\} \, G^{R}(e,\mu,\dots) &=& 0,
\label{Can2}
\end{eqnarray}
where for definiteness we use the time--coordinate representation and
denote ${\cal D}_{\bot} = x_{\bot} \partial / \partial x_{\bot}$,
${\cal D}_{\parallel} = x_{\parallel} \partial / \partial x_{\parallel}$.

The equations of the type (\ref{RGFP}), (\ref{Canon}) and (\ref{Can2})
describe the scaling behaviour of the function $G^{R}$ upon the dilation
of a part of its parameters: a parameter is dilated if the corresponding
derivative enters the equation; otherwise it is kept fixed. We are interested
in the IR scaling behaviour, in which all the IR relevant parameters
(coordinates, times, integral scale) are dilated, while the irrelevant
parameters (diffusivity coefficients, coupling constant) are fixed.
Thus we combine the equations (\ref{RGFP}) and (\ref{Can2}) so that the
derivatives with respect to the IR irrelevant parameters be eliminated;
this gives the desired equation which describes the IR scaling behaviour:
\begin{eqnarray}
\left\{ {\cal D}_{\bot} + \Delta_{\parallel} {\cal D}_{\parallel} +
\Delta_{\omega} {\cal D}_{\omega} + \Delta_{m} {\cal D}_{m} -
\Delta_{G} \right\} G^{R} =0 .
\label{Krit}
\end{eqnarray}
Here $\Delta_{\bot}=1$ is the normalization condition, while the critical
dimensions of any other IR relevant parameter $F$ is given by the general
expression
\begin{eqnarray}
\Delta_{F} = d_{F}^{\bot}+ \Delta_{\parallel} d_{F}^{\parallel}+
\Delta_{\omega} d_{F}^{\omega} + \gamma^{*}_{F},
\label{KritDim}
\end{eqnarray}
where
\begin{eqnarray}
\Delta_{\omega}=2-\gamma_{\nu}^{*} , \quad \Delta_{\parallel} =
\left( 2+\gamma_{f}^{*} \right)/2.
\label{KritDim2}
\end{eqnarray}
Then using (\ref{Nu}), (\ref{Anom}) and the data from table~\ref{table1}
we obtain the following exact expressions for the dimensions:
\begin{eqnarray}
\Delta_{\omega}=2, \ \Delta_{\parallel} = 1+\xi/2, \
\Delta_{\theta'}= d+1+\xi/2, \ \Delta_{\theta}=-1, \ \Delta_{m}=1.
\label{KDE}
\end{eqnarray}

Solution of the RG equations for the structure functions (\ref{struc})
will be discussed in section~\ref{sec:IRA}.

\section{Critical dimensions of the composite operators}
\label{sec:Operators}

The key role in the following will be played by the critical dimensions
of certain composite fields (``composite operators'' in quantum-field
terminology). Detailed exposition of the renormalization procedure of
composite operators can be found in \cite{Book3}. In general, counterterms
to a given operator $F$ are determined by all possible 1-irreducible Green
functions with one operator $F$ and arbitrary number of primary fields,
$\Gamma=\langle F(x) \Phi(x_{1})\dots\Phi(x_{2})\rangle_{\rm 1-ir}$.
The total canonical dimension of this function
(formal index of divergence) is
\begin{equation}
d_\Gamma = d_{F} - \sum_{\Phi} N_{\Phi} d_{\Phi},
\label{index}
\end{equation}
where $d_{F}$ is the dimension of the operator and the summation over
all types of fields is implied, cf. expression (\ref{dGamma}).
For superficially divergent diagrams $d_\Gamma$ is a nonnegative integer.

We begin with the simplest operators $\theta^{n}(x)$, which enter
the structure functions (\ref{struc}). Using the data in table \ref{table1}
we obtain $d_{F}=nd_{\theta}= -n$ and
$d_\Gamma = -n + N_{\theta}-N_{v} -(d+1)N_{\theta'}$.
From the analysis of the diagrams it follows that the total number of the
fields $\theta$ entering the function $\Gamma$ cannot exceed the number of
the fields $\theta$ in the operator $\theta^{n}$ itself, that is,
$N_{\theta}\le n$. This is a direct consequence of the linearity of the
original stochastic equation (\ref{eq1}) in the field $\theta$: the solid
lines in the diagrams cannot branch. Therefore, the superficial divergence
can only exist in the functions with $N_{v}= N_{\theta'}=0$ and arbitrary
value of $n=N_{\theta}$, for which the formal index vanishes, $d_\Gamma =0$.
However, in any nontrivial diagram at least one of $N_{\theta}$ external
`tails' of the field $\theta$ is attached to a vertex
$\theta'(v\partial)\theta$, at least one derivative $\partial$ appears
as an extra factor in the diagram, and, consequently, the real index of
divergence $d_\Gamma'$ is necessarily negative.

This means that the operator $\theta^{n}$ is in fact UV finite and requires
no counterterms at all: $\theta^{n}=Z\,[\theta^{n}]_{R}$ with $Z=1$.
It then follows that its critical dimension, given by the
expression (\ref{KritDim}) without the contribution of $\gamma_{F}^{*}$,
is simply given by the sum of the critical dimensions of its constituents:
\begin{equation}
\Delta [\theta^{n}] = n\Delta_{\theta} = -n.
\label{2.6}
\end{equation}

In Kraichnan's rapid-change model, the anomalous exponents (\ref{HZ3})
are identified with the critical dimensions of the scalar composite
operators of the form $(\partial\theta)^{2n}$, the operators with minimal
canonical dimension (namely, $d_{F}=0$) that are invariant with respect to
the shift $\theta\to\theta+{\rm const}$; see \cite{RG}--\cite{JphysA}. For
that isotropic case, the scalar operator of the needed form is unique
for any given $n$: $F_{n}=(\partial_{i}\theta \partial_{i}\theta)^{n}$.
The operator $F_{n}$ does not admix in renormalization to $F_{k}$ with
$k<n$, the corresponding renormalization matrix is triangular, and the
dimensions $\Delta_{n}$ are determined by its diagonal elements.
(The operators $F_{n}$ can be treated as if they were multiplicatively
renormalizable.) For the existence of the singular behaviour
of the structure functions (\ref{HZ1}) at $(r/{\cal L})\to0$
it is crucial that the dimensions $\Delta_{n}$ are negative.

In our case the $O_{d}$ isotropy is broken to $O_{d-1}\otimes Z_{2}$, and
for any $n$ one can construct a set of $(n+1)$ different operators of the
form $(\partial\theta)^{2n}$, invariant under the residual symmetry, namely:
\begin{equation}
F_{k,s} = (\partial_{i}^{\bot}\theta \partial_{i}^{\bot}\theta)^{k}
(\partial_{\parallel}\theta \partial_{\parallel}\theta)^{s}, \quad k+s=n.
\label{operators}
\end{equation}
Here $\partial_{i}^{\bot}$ is the derivative in the ${\bf x}_{\bot}$
subspace, $\partial_{\parallel} = \partial/ \partial x_{\parallel}$
and the summation over the vector index $i=1,\dots,d-1$ is implied.
In particular, for $n=1$ such a set includes two operators
\begin{equation}
F_{1,0} = (\partial_{i}^{\bot}\theta \partial_{i}^{\bot}\theta),
\quad
F_{0,1} = (\partial_{\parallel}\theta)^{2},
\label{oper1}
\end{equation}
for $n=2$ --- three operators
\begin{equation}
F_{2,0} = (\partial_{i}^{\bot}\theta
\partial_{i}^{\bot}\theta)^{2}, \quad F_{1,1} =
(\partial_{i}^{\bot}\theta \partial_{i}^{\bot}\theta)
(\partial_{\parallel}\theta)^{2}, \quad F_{0,2} =
(\partial_{\parallel}\theta)^{4},
\label{oper2}
\end{equation}
and so on.

For all operators (\ref{operators}), from table~\ref{table1} and
equation (\ref{index}) we find $d_{F}=0$ and
$d_\Gamma = N_{\theta}-N_{v} -(d+1)N_{\theta'}$, with the necessary
condition $N_{\theta}\le 2n=2(k+s)$, following from the linearity of the
model (cf. the discussion for $\theta^{n}$ above). From the form of the
vertex $\theta'(v\partial)\theta$ and the operators $F_{k,s}$ themselves
it follows that the fields $\theta$, $\theta'$ enter the counterterms
only in the form of derivatives, $\partial\theta$, $\partial\theta'$,
so that the real index of divergence is
$d_\Gamma' = d_\Gamma -N_{\theta}-N_{\theta'}=-N_{v} -(d+2)N_{\theta'}$.
Thus the superficial divergences can only exist in the Green functions
with $N_{v}=N_{\theta'}=0$ and arbitrary $N_{\theta}\le 2n$
(then $d_\Gamma' =0$), and the corresponding operator counterterm
necessarily reduces to the form $F_{m,l}$ with $(m+l)\le (k+s)$.

Therefore, the operators of the type $F_{k,s}$ can mix only with each
other in renormalization, the corresponding infinite renormalization matrix
$Z=\{Z_{k,s;m,l}\}$ is in fact block-triangular, $Z_{k,s;m,l}=0$ for
$(m+l)>(k+s)$, and the critical dimensions associated with the set of
operators with given $n=(k+s)$ are determined by its finite diagonal blocks
$Z_{n}=\{Z_{k,s;m,l}\}$ with the fixed value of the sum $n=(m+l)=(k+s)$.

In the following, we will show that the counterterms and the renormalization
matrices for these operators are fully specified by the one-loop
approximations.

\begin{figure}
\begin{center}
\includegraphics[width=12cm]{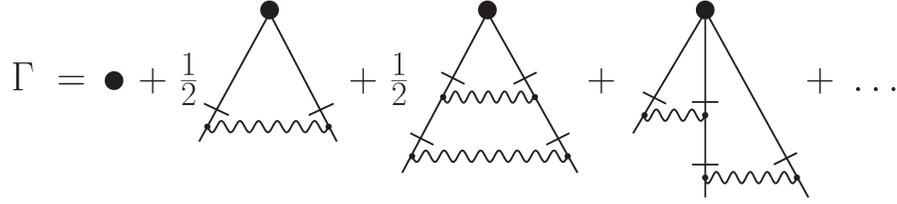}
\caption{\label{fig:oper}
The one-loop and a part of the two-loop contributions to
the generating functional (\protect\ref{GrandFunk}).}
\end{center}
\end{figure}

It is convenient to work with the generating functional of 1-irreducible
correlation functions with one operator $F_{k,s}$ and arbitrary number
of the primary fields $\theta$:
\begin{eqnarray}
\Gamma_{k,s}(x;\theta) = \sum_{p=0}^{\infty} \int dx_{1} \dots \int dx_{p}
\bigg\langle F_{k,s}(x) \theta (x_{1})\cdots \theta (x_{p})
\bigg\rangle_{\rm 1-ir} \theta (x_{1})\cdots \theta (x_{p}).
\nonumber \\
{} \label{GrandFunk}
\end{eqnarray}
In the diagrammatic expansion of the functional $\Gamma_{k,s}$, the leading
(loopless) approximation is given by the operator $F_{k,s}(x)$ itself.
The only one-loop diagram and two of the five two-loop diagrams are shown
in figure~\ref{fig:oper}. The diagrammatic notation was explained in
section~\ref{sec:Reno} in connection with figure~\ref{fig:sigma}; the new
elements are the thick dots that denote the vertex factors of the composite
field $F_{k,s}$:
\begin{equation}
V^{(l)}_{k,s}(x;x_{1},\dots, x_{l}) =
\frac{\delta^{l}F_{k,s}(x)}{\delta\theta(x_{1})\cdots\delta\theta(x_{l})},
\label{VerF}
\end{equation}
where $l$ is the number of $\langle\theta\theta'\rangle_{0}$-lines attached
to the vertex (\ref{VerF}). The variational derivative acts on such
operators as follows:
\begin{eqnarray}
\frac{\delta}{\delta\theta(x')} F_{k,s}(x) &=& 2k\, F_{k-1,s}(x)\,
\partial_{i}^{\bot}\theta(x)\, \partial_{i}^{\bot} \delta(x-x') +
\nonumber \\
&+& 2s\, F_{k,s-1}(x)\, \partial_{\parallel}\theta(x)
\partial_{\parallel} \delta(x-x'),
\label{VarD}
\end{eqnarray}
so that the vertex factor (\ref{VerF}) is in fact local.

The derivatives $\partial_{\parallel}$ at the external vertices (two lower
vertices in the first two diagrams on figure~\ref{fig:oper} and three lower
vertices in the last diagram) correspond to external momenta; they act onto
the fields $\theta$ attached to those vertices and form the external factors
$\sim (\partial_{\parallel}\theta)^{l}$. The remaining expressions diverge
logarithmically and therefore can be calculated at all external momenta equal
to zero; the IR regularization is provided by the cutoff $m$ in the velocity
correlator (\ref{veloc2}).

It is easy to see that the number of independent loops in any diagram of
the functional (\ref{GrandFunk}) equals to the number of wavy lines; thus
the independent integration momenta (denoted as ${\bf k}$ and ${\bf q}$
below) can always be assigned to the velocity correlators. Then the integrals
corresponding to the diagrams in figure~\ref{fig:oper} take on the forms:
\begin{equation}
\int\frac{d{\bf k}}{(2\pi)^{d}} \int\frac{d\omega}{2\pi}
\frac{k_{i}k_{j}} {(\omega^{2}+ \epsilon^{2}({\bf k}) )}
\left[ D_{v} (k) (\partial_{\parallel}\theta(x))^{2} \right]
\label{Old}
\end{equation}
for the one-loop diagram,
\begin{eqnarray}
\int\frac{d{\bf k}}{(2\pi)^{d}}\int\frac{d{\bf q}}{(2\pi)^{d}}
\int\frac{d\omega}{2\pi}\int\frac{d\Omega}{2\pi} \,
\left[ D_{v} (k) q^{2}_{\parallel} \right]\left[ D_{v} (q)
(\partial_{\parallel}\theta(x))^{2} \right]
\times \nonumber \\ \times
\frac{(k+q)_{i}(k+q)_{j}} { \left\{\omega^{2}+ \epsilon^{2}({\bf k}+{\bf q})
+ \epsilon^{2}({\bf q})\right\}
\left\{ \Omega^{2}+\epsilon^{2}({\bf q})\right\} }
\label{Tld}
\end{eqnarray}
for the first two-loop diagram and
\begin{eqnarray}
\int\frac{d{\bf k}}{(2\pi)^{d}}\int\frac{d{\bf q}}{(2\pi)^{d}}
\int\frac{d\omega}{2\pi}\int\frac{d\Omega}{2\pi}\,
\left[ D_{v} (k) ({\rm i} q_{\parallel} \partial_{\parallel}
\theta(x))\right]
\left[ D_{v} (q) (\partial_{\parallel}\theta(x))^{2} \right]
\times \nonumber \\ \times
\frac{{\rm i}k_{i}q_{j}(k+q)_{l}} {
\left\{ \Omega^{2}+\epsilon^{2} ({\bf q}) \right\}
\left\{-{\rm i}\omega+\epsilon({\bf k}) \right\}
\left\{{\rm i}(\omega+\Omega)+  \epsilon({\bf k}+{\bf q})\right\} }
\label{Tld2}
\end{eqnarray}
for the last one. Here
$\epsilon({\bf k})=\nu (k_{\bot}^{2}+fk_{\parallel}^{2})$
is the renormalized analog of the quantity from (\ref{lines2}).

The momenta with free vector indices stem from the
vertex factor (\ref{VerF}) and correspond to the derivatives acting onto
the spatial $\delta$ functions in expressions like (\ref{VarD}). Each factor
in square brackets comes from one velocity correlator (\ref{veloc1}) and two
vertices (\ref{vertex}) attached to it; $D_{v}$ being the function
(\ref{veloc2}). Thus the integrand for any diagram contains the factor
$D_{v} (k) \sim \delta(k_{\parallel})$ for each independent integration
momentum ${\bf k}$. On the other hand, any diagram of the functional
(\ref{GrandFunk}) with more than one loop necessarily contains at least one
internal vertex (\ref{vertex}), the corresponding vertex factor is
proportional to certain linear combination of independent parallel momenta
$k_{\parallel}$, $q_{\parallel}$ and so on, and the integrand as a whole
vanishes due to the presence of the corresponding $\delta$ functions
(we recall that all the external momenta are set equal to zero). Thus the
only contribution to the counterterm comes from the one-loop diagram. For
the same reason, the integral (\ref{Old}) is nontrivial only if the both
momenta in the factor $k_{i}k_{j}$ are perpendicular to the vector ${\bf n}$;
otherwise $k_{i}= n_{i} k_{\parallel}$ and the above mechanism works. So
one can replace $k_{i}\to k_{i}^{\bot}$ in (\ref{Old}), and in the vertex
keep only the term
\begin{equation}
V^{(2)}_{k,s}(x;x_{1},x_{2}) \simeq F_{0,s}(x)
\frac{\delta^{2}}{\delta\theta(x_{1})\delta\theta(x_{2})} F_{k,0}(x).
\label{Ver1}
\end{equation}
Thus the operator $F_{k,s}$ in the divergent part
of the one-loop diagram loses
two factors $\partial^{\bot}\theta$ in (\ref{Ver1}) and acquires two factors
$\partial_{\parallel}\theta$ in (\ref{Old}), so that the counterterm to the
functional (\ref{GrandFunk}) reduces to the form $F_{k-1,s+1}$. We conclude
that the renormalized operator corresponding to $F_{k,s}$ has the form
$F^{R}_{k,s}= F_{k,s} +a F_{k-1,s+1} $. Resolving these relations for
$F_{k,s}$ gives
\begin{equation}
F_{k,s}= F^{R}_{k,s} +\sum_{l=1}^{l=k} a_{l} F^{R}_{k-l,s+l}
\label{RF}
\end{equation}
with some coefficients $a_{l}$ and $a_{1}=-a$. Thus the renormalization
matrix $Z_{n}=\{Z_{k,s;m,l}\}$ in the matrix relation $F=Z_{n} F^{R}$
is triangular (with a natural numbering of operators in the family with
fixed $k+s=n$) and its diagonal elements are all equal to unity,
$Z_{k,s;k,s}=1$. The other nonvanishing elements $Z_{k,s;k-l,s+l}$ are
determined by the one-loop diagram (\ref{Old}), but we will not need their
explicit expressions. Indeed, the matrix of critical dimensions for the
operators (\ref{operators}) is given by the expressions
(\ref{KritDim})--(\ref{KDE}), in which $d_{F}^{*}$ should be understood
as diagonal matrices of canonical dimensions of the operators $F_{k,s}$
and $\gamma_{F}=Z_{n}^{-1}\widetilde{\cal D}_{\mu} Z_{n}$ is the matrix
of their anomalous dimensions. The latter is triangular with vanishing
diagonal elements $\gamma_{k,s;k,s}=0$ and nontrivial $\gamma_{k,s;k-l,s+l}$
with $k=1,\dots,l$. Thus the critical dimensions $\Delta_{k,s}$ related
to the set $F_{k,s}$, given by the eigenvalues of the matrix
(\ref{KritDim}), coincide with its diagonal elements.
From table~\ref{table1} for the operators $F_{k,s}$ we find
\[ d_{F}^{\bot} = 2k, \quad d_{F}^{\parallel} = 2s,
\quad  d_{F}^{\omega} = - (k+s),  \]
so that
\[ \Delta_{k,s} = 2k + 2s \Delta_{\parallel} - (k+s) \Delta_{\omega} \]
with no contribution from $\gamma^{*}_{F}$, which along with expressions
(\ref{KDE}) gives the final exact result
\begin{equation}
\Delta_{k,s} = s \xi
\label{KDF}
\end{equation}
for the critical dimensions related to the operators (\ref{operators}).
In contrast to the result (\ref{HZ3}) for the isotropic Kraichnan's
model, they have no corrections of order $O(\xi^{2})$ and higher and are
positive for all $k$, $s$ and $\xi>0$.

\section{RG and the IR scaling behaviour of the structure functions}
\label{sec:IRA}

The results of the preceding sections allow one to find the IR scaling
behaviour of the correlation functions in our model. For generality,
consider the different-time structure functions
\begin{equation}
S_{2n}(\tau, r_{\bot}, r_{\parallel} ) =
\langle\left[\theta(t,{\bf x})-\theta(t',{\bf x}')\right]^{2n} \rangle,
\label{struT}
\end{equation}
where $\tau=t'-t$, $r_{\bot} =|{\bf x}_{\bot}-{\bf x}'_{\bot}|$ and
$r_{\parallel} = |x_{\parallel}-x_{\parallel}'|$. The equal-time functions
(\ref{struc}) are obtained for $\tau=0$.

The function (\ref{struT}) is a linear combination of the two-point
correlators $\langle \theta^{k} (t,{\bf x})\theta^{s} (t',{\bf x}')\rangle$
with the fixed $k+s=2n$. Due to simple exact relations (\ref{2.6}) for the
dimensions of the operators $\theta^{n}$, the critical dimensions of these
correlators are all equal,  $\Delta_{k}+\Delta_{s} = -(k+s) = -2n =
2n \Delta_{\theta}$, and the function $S_{2n}$ in the IR range behaves
as a single object. Its IR scaling form can be easily written using the
IR scaling equation (\ref{Krit}) with
known critical dimensions (\ref{KDE}) of the basic IR relevant parameters,
but it is instructive to discuss its derivation from the RG equation in
more detail.

From the dimensional considerations one can write
\begin{equation}
S_{2n}= \nu^{-n} r_{\bot}^{2n} {\cal F} \left( \mu r_{\bot}, w,
r_{\parallel} / f^{1/2}r_{\bot}, \nu \tau r_{\bot}^{2},
mr_{\bot} \right),
\label{struD}
\end{equation}
where ${\cal F}(\dots)$ are some functions of completely dimensionless
arguments (that is, dimensionless with respect to all the canonical
dimensions $d_{F}^{*}$ separately; see section~\ref{sec:Reno}).

Since the operators $\theta^{n}$ are not renormalized, the function $S_{2n}$
satisfies the homogeneous RG equation (\ref{RG1}) with $\gamma_{G}=0$.
Its solution can be represented in the form
\begin{equation}
S_{2n}= \bar\nu^{-n} r_{\bot}^{2n} {\cal F} \left( 1, \bar w,
r_{\parallel} / \bar f^{1/2}r_{\bot}, \bar\nu \tau/ r_{\bot}^{2},
\bar mr_{\bot} \right),
\label{Solu}
\end{equation}
with the same functions ${\cal F}$. The invariant variables $\bar e$ are
solutions of the homogeneous equation (\ref{RG1}) normalized with respect
to $e$ at $\mu r_{\bot}=1$. They can be expressed in the original (bare)
parameters $e_{0}$ using the relations
\begin{equation}
w_{0} = \bar w r_{\bot}^{-\xi}  Z_{w}(\bar w), \quad
f_{0} = \bar f Z_{f} (\bar w), \quad
\bar\nu=\nu_{0}, \quad \bar m= m_{0},
\label{invar}
\end{equation}
with the renormalization constants from (\ref{18}) and (\ref{NoRen}).
The representations (\ref{Solu}) and (\ref{invar}) are valid because
both their sides satisfy the same RG equation and coincide in the
normalization point $\mu r_{\bot}=1$ due to the definition of $\bar e$
and the relations (\ref{18}). The simple result for $\bar\nu$ and
$\bar m$ follows from the fact that these parameters are not renormalized.

The advantage of the representation (\ref{Solu}) is that the invariant
variables have simple asymptotic behaviour for $\Lambda r_{\bot}\to\infty$
(or $\mu r_{\bot} \to \infty$ in renormalized variables).
The exact explicit expression for the invariant charge
\begin{equation}
\bar w = w_{*} \left[ 1+ w_{*} r_{\bot}^{-\xi}/w_{0}  \right]^{-1}
\label{InC}
\end{equation}
is derived from the relations (\ref{18}), (\ref{Zf}) and (\ref{invar}).
From (\ref{InC}) it follows that, for $\xi>0$ and $r_{\bot} \to \infty$,
the invariant charge tends to the IR attractive fixed point of the RG
equation: $\bar w \to w_{*}=\xi$. For $\bar f$, eliminating
the renormalization constant $Z_{w}=Z_{f}^{-1}$ from (\ref{invar}) gives
\begin{equation}
\bar f = f_{0}w_{0} r_{\bot}^{\xi} / \bar w \to
f_{0}w_{0} r_{\bot}^{\xi} / w_{*}.
\label{InF}
\end{equation}
Substituting these expressions into (\ref{Solu}) gives the desired
asymptotic expression for the structure functions in the IR range
$\Lambda r_{\bot} \gg 1$:
\begin{equation}
S_{2n}= \nu_{0}^{-n} r_{\bot}^{2n} {\cal F}
\left( 1, w_{*}, r_{\parallel} / (f_{0}w_{0})^{1/2} r_{\bot}^{1+\xi/2},
\nu_{0} \tau/ r_{\bot}^{2}, mr_{\bot} \right) .
\label{AsSou}
\end{equation}
It can be made more transparent by discarding in the notation all the IR
irrelevant parameters:
\begin{eqnarray}
S_{2n} &=&
r_{\bot}^{2n} {\cal R} \left( r_{\parallel} r_{\bot}^{-1-\xi/2},
\tau/ r_{\bot}^{2}, mr_{\bot} \right)
= \nonumber \\ &=&
r_{\bot}^{-2n\Delta_{\theta}} {\cal R} \left( r_{\parallel} /
r_{\bot}^{\Delta_{\parallel}},  \tau / r_{\bot}^{\Delta_{\omega}},
m r_{\bot}^{\Delta_{m}} \right)
\label{ScS}
\end{eqnarray}
with certain scaling functions ${\cal R}$. Of course, the last expression,
in which we have used the explicit forms (\ref{KDE}) of the critical
dimensions, could be derived directly from the general IR scaling equation
(\ref{Krit}), but the more detailed representation (\ref{AsSou}) gives
additional interesting information about the dependence on the IR irrelevant
papameters.

In the isotropic Kraichnan's model, the structure functions in the IR range
$\Lambda r_{\bot}\gg 1$ depend only on the IR (integral) scale ${\cal L}=1/m$
and the amplitude $D_{0}=w_{0}\nu_{0}$ entering the velocity correlator
(\ref{OK}), but not on the diffusivity coefficient $\nu_{0}$ and the coupling
constant $w_{0} \sim \Lambda^{\xi}$ separately; see (\ref{HZ1}). The same
effect takes place for the stochastic Navier--Stokes equation; see e.g. the
discussion in chapter~6 of \cite{Book3}. This fact is in agreement with the
second Kolmogorov hypothesis about the independence of the correlation
functions in the IR range of the parameters, related to the UV (dissipation)
scale: viscosity coefficient for the velocity and diffusivity coefficient
for the passive scalar field; see e.g. \cite{Legacy}.
In the case at hand, the UV parameters $\nu_{0}$, $f_{0}$ and
$w_{0} \sim \ell^{-\xi}$ survive in the IR asymptotic expression
(\ref{AsSou}) for the structure functions (they do not form the
combination $D_{0} = w_{0}\nu_0 f_{0}$ even if we set $f_{0}=1$).
Thus we may conclude that, in contrast to the isotropic case,
the second Kolmogorov hypothesis is invalid for the shear flow.

\section{OPE and the inertial-range behaviour of the structure functions}
\label{sec:OPE}

The asymptotic representations (\ref{AsSou}) and (\ref{ScS}) hold in the
IR asymtotic range, specified by the inequality $\Lambda r_{\bot} \gg 1$
(or $\mu r_{\bot} \gg 1$ in renormalized variables), while
the other arguments of the scaling functions ${\cal R}$ are kept finite.
Inertial range corresponds to the additional condition $mr_{\bot} \ll 1$.
The form of the scaling functions ${\cal F}$ or ${\cal R}$ is not
determined by the RG equations alone. In order to study the limit
$mr_{\bot}\to0$ in the structure functions, one should combine the plain RG
with the OPE techniques; see \cite{RG}--\cite{JphysA} for Kraichnan's model.
For simplicity, below we concentrate on the equal-time functions, setting
$\tau=0$ in (\ref{AsSou}) and (\ref{ScS}).

According to the OPE, the behaviour of the quantities entering into
the structure functions (\ref{struc}) for
${\bf r} = {\bf x} - {\bf x'} \to 0$  and fixed
$ {\bf x} + {\bf x'} $ is given by the infinite sum
\begin{equation}
\left[ \theta (t,{\bf x}) - \theta (t,{\bf x'})\right]^{n}
= \sum_{F} C_{F} ({\bf r}) F\left(t,\, \frac{{\bf x}+{\bf x'}}{2} \right),
\label{OPE}
\end{equation}
where $C_{F}$ are coefficients regular in $m^{2}$ and $F$ are all possible
renormalized local composite operators allowed by the symmetry. More
precisely, the operators entering the OPE are those which appear in the
plain Taylor expansion, and all the operators that admix to them
in renormalization.

The structure functions (\ref{struc}) are obtained by averaging equation
(\ref{OPE}) with the weight $\exp {\cal S}_{R}$, where ${\cal S}_{R}$ is
the renormalized action; the mean values
$\langle F\rangle$ appear on the right-hand side. Their asymptotic
behaviour for $m\to0$ is found from the corresponding RG equations
and has the form $\langle F\rangle \propto  m^{\Delta_{F}}$. Then
substituting the OPE into the asymptotic expression (\ref{ScS}) gives
\begin{equation}
S_{2n} ({\bf r})= r_{\bot}^{2n} \sum_{F} A_{F}
\left(r_{\parallel} / r_{\bot}^{\Delta_{\parallel}}, m r_{\bot} \right)
\, (mr_{\bot})^{\Delta_{F}} ,
\label{OR}
\end{equation}
with coefficients $A_{F}$ regular in $(mr_{\bot})^{2}$.

Due to the linearity of the model, the number of the fields $\theta$ in
such operators $F$ cannot exceed their number in the
left-hand side. Due to the invariance of the action functional
(\ref{action}) and the quantity in (\ref{OPE}) with respect to the
shift of the field, $\theta\to\theta+{\rm const}$, the operators entering
the OPE must also obey this symmetry.
It can always be assumed that the expansion (\ref{OPE}) is made in
irreducible tensors (scalars, vectors and traceless tensors); owing to
the symmetries of our model, only contributions of the scalar operators
survive in (\ref{OR}). The leading contributions are determined by the
operators with minimal critical dimensions $\Delta_{F}=d_{F}+ O(\xi)$.

It then follows that the leading terms of the small-$mr_{\bot}$ behaviour of
the expression (\ref{OR}) for the function $S_{2n}$ are given by the scalar
operators $F_{k,s}$ from (\ref{operators}) with $k+s \le n$.
Their dimensions $\Delta_{k,s} = s \xi$ in (\ref{KDF}) are all nonnegative;
the leading term is given by the operators with $s=0$ (including the
simplest $F=1$), the operators with $s \ge 1$ determine the corrections
vanishing for $mr_{\bot}\to0$.

We conclude that the function $S_{2n}$ remains finite at $m=0$. Thus in the
inertial range and for $\tau=0$ expression (\ref{ScS}) becomes
\begin{eqnarray}
S_{2n} = r_{\bot}^{-2n\Delta_{\theta}} {\cal R} \left( r_{\parallel} /
r_{\bot}^{\Delta_{\parallel}}  \right),
\label{ScC}
\end{eqnarray}
which for $r_{\parallel}=0$ turns to the simple power law
\begin{equation}
 S_{2n} \sim r_{\bot}^{-2n\Delta_{\theta}} = r_{\bot}^{2n},
\label{A1}
\end{equation}
while for $r_{\bot}=0$ one obtains
\begin{equation}
S_{2n} \sim r_{\parallel}^{-2n\Delta_{\theta}/\Delta_{\parallel}}
=  r_{\parallel}^{2n/(1+\xi/2)}.
\label{A2}
\end{equation}

Thus the inertial-range behaviour in the rapid-change model of a shear
flow differs essentially from that in the isotropic model. In contrast
to the behaviour (\ref{HZ1}), (\ref{HZ3}) of the latter,  singular at
$mr\to0$, there is no anomalous (multi)scaling in the case at hand.

\section{Finite correlation time} \label{sec:FCT}

In this section we will consider the case of the Gaussian velocity field
with a finite correlation time. Now the function $\widetilde D_{v}$
in the correlator (\ref{veloc1}), (\ref{veloc2}) depends on frequency
$\omega$ and will be chosen in the form
\begin{eqnarray}
\widetilde D_{v} (\omega, k_{\bot}) = \frac{g_{0} \nu_0^{3} f_{0}\,
k_{\bot}^{5-d-(\varepsilon+\eta)}} {\omega^{2} +
[u_{0} \nu_0 k_{\bot}^{2-\eta}]^{2} }.
\label{velocF}
\end{eqnarray}
The function (\ref{velocF}) involves two independent exponents $\eta$
and $\varepsilon$, which in the RG approach play the role of two formal
expansion parameters. The former defines the dispersion law
$\omega(k_{\bot}) \sim k_{\bot}^{2-\eta}$ ($z=2-\eta$ in the notation
of \cite{AM}--\cite{Walls}), while the latter governs the behaviour of the
one-dimensional velocity spectrum, related to the equal-time correlator:
\begin{eqnarray}
{\cal E} (k_{\bot}) \sim k_{\bot}^{d-2} \int \frac{d\omega}{2\pi}
\widetilde D_{v} (\omega, k_{\bot}) =
(g_{0} f_{0} \nu_0^{2} /2u_{0}) k_{\bot}^{1-\varepsilon} ;
\label{spektr}
\end{eqnarray}
this explains the choice of the exponent in the numerator of (\ref{velocF}).

The correlator (\ref{velocF}) involves two important special cases, which,
as we will see, nearly exhaust possible IR behaviour of the model.
The limit $u_{0}\to 0$ at fixed $g_{0}' = g_{0}/u_{0}$ corresponds to
the case of time-independent (`frozen') velocity field, when (\ref{velocF})
turns to $\widetilde D_{v} \sim \delta(\omega) k^{3-d-\varepsilon}$.
The limit $u_{0}\to \infty$ at fixed $g_{0}'' = g_{0}/u^{2}_{0}$ returns
us to the rapid-change case (\ref{veloc1}), (\ref{veloc2}) with
$\xi = \varepsilon - \eta$ and $w_{0}=g_{0}''$.

The role of coupling constants will be played by the parameters (see below)
\begin{eqnarray}
g_{0} \sim \Lambda^{\varepsilon+\eta} \sim \Lambda^{\xi+2\eta},
\quad u_{0} \sim \Lambda^{\eta}, \quad g_{0}' \sim \Lambda^{\varepsilon},
\quad g_{0}'' \sim \Lambda^{\xi},
\label{kopls}
\end{eqnarray}
where $\Lambda$ is a typical UV momentum scale, cf.~(\ref{g0}).
The model is logarithmic (the couplings in (\ref{kopls}) are all
dimensionless) at $\eta=\varepsilon=0$, so that the UV divergences manifest
themselves as poles in $\eta$, $\varepsilon$ and their linear combinations.

As a rule, synthetic velocity ensembles with a finite correlation time
suffer from the lack of Galilean invariance, which can lead to some
physical pathologies; see e.g. the discussion in \cite{synth}.
Surprisingly enough, in our strongly anisotropic case the action functional
(\ref{action}) with the correlator (\ref{velocF})  in (\ref{Sv}) is
invariant with respect to the Galilean transformation of the fields
\begin{eqnarray}
\theta(t,{\bf x}) \to \theta(t,{\bf x}+{\bf u}t), \quad
\theta'(t,{\bf x}) \to \theta'(t,{\bf x}+{\bf u}t), \nonumber \\
{\bf v}(t, {\bf x}) \to {\bf v}(t, {\bf x} +{\bf u}t) - {\bf u},
\label{GT}
\end{eqnarray}
where the transformation parameter has the form ${\bf u}={\bf n} u$ with
vector ${\bf n}$ from (\ref{vello}), so that the scalar coefficient in
(\ref{vello}) changes as $v(t, {\bf x}_{\bot}) \to v(t, {\bf x}_{\bot})-u$
and the arguments ${\bf x}_{\bot}$ of all the fields in (\ref{GT}) remain
intact. This fact can be interpreted as follows. Consider the stochastic
Navier--Stokes equation
\begin{equation}
\partial_{t} v_{i} + (v_{l} \partial_{l}) v_{i} + \partial_{i} \wp
= R(\partial) v_{i} + \phi_{i},
\label{NS}
\end{equation}
where $R(\partial)$ is some linear differential operation, $\wp= -
\partial^{-2} (\partial_{i}v_{l})(\partial_{l} v_{i})$ is the pressure
and $\phi_{i}$ is a white-in-time random force. The equation (\ref{NS})
is of course Galilean covariant. For the velocity field of the form
(\ref{vello}) nonlinear terms in (\ref{NS}) vanish due to the independence
of the scalar coefficient $v$ on $x_{\parallel}$:
$ v_{k}\partial_{k} v_{i} = n_{i} v \partial_{\parallel} v =0$
and similarly for the pressure. Thus the equation (\ref{NS}) becomes
in fact linear and generates a Gaussian velocity field.
Its correlator coincides with (\ref{velocF}) if one choses
$R(k) = u_{0} \nu_0 k_{\bot}^{2-\eta}$ and $\phi_{i} = \phi n_{i}$
with $\langle \phi\phi \rangle = g_{0} \nu_0^{3} f_{0}\, \delta(t-t')\,
\delta(k_{\parallel}) k_{\bot}^{5-d-(\varepsilon+\eta)}$.

The analysis similar to that performed in section~\ref{sec:Reno} shows that
the UV divergences in the model are removed by the only counterterm
$\theta' \partial_{\parallel}^{2}  \theta$. Thus the model is
multiplicatively renormalizable with the only independent
renormalization constant $Z_{f}$:
\begin{eqnarray}
g_{0}=g\mu ^{\varepsilon+\eta} Z_{g}, \quad u_{0}=u\mu ^{\eta} Z_{u},
\quad f_{0}=f Z_{f}, \quad \nu_{0}= \nu Z_{\nu},
\label{RenoF}
\end{eqnarray}
where
\begin{equation}
Z_{g}= Z_{f}^{-1}, \quad Z_{u}=Z_{\nu}=1.
\label{ReCF}
\end{equation}

The constant $Z_{f}$ is found from the requirement that the 1-irreducible
Green function $\langle\theta'\theta\rangle_{\rm 1-ir}$, expressed in
renormalized variables, be finite at $\varepsilon=\eta=0$. Since the
counterterm has the form $\theta' \partial_{\parallel}^{2}  \theta$, the
UV divergent part of the self-energy operator $\Sigma(\omega, {\bf p})$
in (\ref{Dy1}) is proportional to $p_{\parallel}^{2}$ and it is
sufficient to calculate it at vanishing external frequency $\omega=0$.
Furthermore, $p_{\parallel}^{2}$ is isolated in any of its diagram
as an extra factor due to the two external vertices (\ref{vertex});
see figure~\ref{fig:sigma} and discussion below equation~(\ref{D2})
in section~\ref{sec:Reno}. Thus in the rest of the corresponding integrals
one can set ${\bf p}=0$, and the mechanism described below equations
(\ref{Old})--(\ref{Tld2}) `kills' all the diagrams with more than one loop.
We stress that, in contrast to the rapid-change case, these diagrams are
nontrivial; it is only their divergent parts that vanish.

We are left with the only one-loop diagram; the corresponding frequency
integral is well-defined, and we obtain
\begin{eqnarray}
\Sigma({\bf p})=  p_{\parallel}^{2} \, \int \frac{d\omega}{2\pi}
\int_{k_{\bot}>m} \frac{d{\bf k}^{\bot}}{(2\pi)^{d-1}} \,
\frac{g\mu^{\varepsilon+\eta} \nu^{3} f\, k_{\bot}^{5-d-(\varepsilon+\eta)}}
{\omega^{2} + [u\mu^{\eta} \nu k_{\bot}^{2-\eta}]^{2} }
\frac{1}{-{\rm i}\omega+\nu k_{\bot}^2 }
= \nonumber \\ =
p_{\parallel}^{2} \, \frac{g\mu^{\varepsilon}\nu^2f}{2u}
\int_{k_{\bot}>m} \frac{d{\bf k}^{\bot}}{(2\pi)^{d-1}} \,
\frac{k_{\bot}^{1-d-{\varepsilon}}} {1+u (\mu/k_{\bot})^{\eta}} .
\label{DF}
\end{eqnarray}
The integral in (\ref{DF}) and thus the renormalization constant $Z_{f}$
can be calculated as expansions in $u$, the individual terms would contain
the poles $\sim 1/(\varepsilon+s\eta)$ with $s\ge 1$;
cf. equations (3.16), (3.17) in \cite{RG3} and (3.18)--(3.20) in \cite{RG4}.
The calculation can be simplified \cite{Juha2} by observing that the
anomalous dimensions in the MS scheme are independent of the UV regulators
like $\varepsilon$ and $\eta$ (at least in the one-loop approximation for
our model with several such regulators), so that it is sufficient to
calculate the constant $Z_{f}$ for $\eta=0$, when the integral (\ref{DF})
remains finite. This gives
\begin{eqnarray}
\Sigma({\bf p})=
p_{\parallel}^{2} \, \frac{g\mu^{\varepsilon}\nu^2f}{2u(u+1)}
\int \frac{d\omega}{2\pi}
\int_{k_{\bot}>m} \frac{d{\bf k}^{\bot}}{(2\pi)^{d-1}} \,
\frac{1}{k_{\bot}^{d-1+{\varepsilon}}}
= \nonumber \\ =
p_{\parallel}^{2} \, \frac{g\nu^2f}{2u(u+1)}
\frac{S_{d-1}}{(2\pi)^{d-1}} \left(\frac{\mu}{m}\right)^{\varepsilon}
\frac{1}{\varepsilon}
\label{D2F}
\end{eqnarray}
and
\begin{equation}
Z_{f} = 1 - \frac{g}{u(u+1)} \frac{1}{\varepsilon} , \quad
\gamma_{f} = \frac{g}{u(u+1)} ,
\label{ZF}
\end{equation}
where the factor ${S_{d-1}}/{2(2\pi)^{d-1}}$ is absorbed into $g$.
We stress that the expression for $\gamma_{f}$ is exact: in contrast to
the isotropic case \cite{Juha2} it has no corrections of order $g^{2}$
and higher.

In general, coordinates $g_{i*}$ of the fixed points in a problem with
several coupling constants are found from the requirement that the
$\beta$-functions, corresponding to all renormalized couplings $g_{i}$,
vanish. The type of a fixed point is determined by the matrix $\Omega$
with the elements $\Omega_{ik}= \partial\beta_{i}/\partial g_{k}$, where
$\beta_{i}$ is the full set of $\beta$-functions and $g_{k}$ is the full
set of couplings. For an IR attractive fixed point the matrix $\Omega$
is positive, that is, the real parts of all its eigenvalues are positive.
In our case $g_{i}= \{g,u\}$: although $u$ is not an expansion parameter
in the perturbation theory, the renormalization constants and anomalous
dimensions depend on it, and it should be treated as an additional coupling
constant. The corresponding functions
$\beta \left(g_{i}\right) = \widetilde {\cal D}_{\mu} g_{i}$
have the forms:
\begin{equation}
\beta_{g}=g[-(\varepsilon+\eta)+\gamma_{f}], \quad
\beta_{u}= -u \eta, \quad  \gamma_{f} = {g}/{u(u+1)}.
\label{betaF}
\end{equation}
Since $\partial \beta_{u}/\partial g =0$, the matrix $\Omega$ is triangular
and its eigenvalues are simply given by the diagonal elements $\Omega_{g}$
and $\Omega_{u}$.

Analysis of the $\beta$ functions (\ref{betaF}) reveals several possible
fixed points:

\bigskip

1) $u_{*}=0$ with $\Omega_{u}=-\eta$. Obviously, this corresponds to the
frozen case (see the remark below equation (\ref{spektr}), and it is
convenient to pass from $g$ to the new coupling $g' = g/u$ with the
$\beta$ function
$\beta_{g'} = \widetilde {\cal D}_{\mu} g' = (1/u) \beta_{g} -
(g/u^{2}) \beta_{u}$, which remains finite at $u=0$:
$\beta_{g'} = g' [-\varepsilon+ \gamma_{f}]= g' [-\varepsilon+ g']$.
Thus we find two fixed points:

\bigskip

1{\it a}) $g'_{*}=0$, $u_{*}=0$ with $\Omega_{g}=-\varepsilon$,
$\Omega_{u}=-\eta$
and $\gamma_{f}^{*}=0$, IR attractive for $\varepsilon<0$, $\eta<0$;

1{\it b}) $g'_{*}=\varepsilon$, $u_{*}=0$ with
$\Omega_{g}= \gamma_{f}^{*}= \varepsilon$, $\Omega_{u}=-\eta$,
IR attractive for $\varepsilon>0$, $\eta<0$.

\bigskip

2) $u_{*}=\infty$. This corresponds to the rapid-change case (\ref{veloc1}),
(\ref{veloc2}), and it is convenient to pass to new charges $v=1/u$,
$g'' = g/u^{2}$ with the $\beta$ functions $\beta_{v}=-(1/u^{2})
\beta_{u} = v\eta$ and
$\beta_{g''} = (1/u^{2}) \beta_{g} - (2g/u^{3}) \beta_{u}
= g'' [-\xi+ \gamma_{f}]$ with $\xi = \varepsilon - \eta$, which for
$v=0$ gives $\beta_{g''} = g'' [-\xi+ g'']$.

\bigskip

Thus we find two more fixed points:

2{\it a}) $g''_{*}=0$, $v_{*}=0$ with $\Omega_{g}=-\xi$, $\Omega_{u}=\eta$
and $\gamma_{f}^{*}=0$, IR attractive for $\xi<0$ (that is,
$\varepsilon<\eta$), $\eta>0$;

2{\it b}) $g''_{*}=\varepsilon$, $v_{*}=0$ with
$\Omega_{g}= \gamma_{f}^{*}= \xi$, $\Omega_{u}=\eta$,
IR attractive for $\xi>0$ (that is, $\varepsilon>\eta$), $\eta>0$.

\bigskip

For the special case $\eta=0$ the function $\beta_{u}$ and the eigenvalue
$\Omega_{u}$ vanish identically, and the nontrivial fixed point, attractive
for $\varepsilon>0$, becomes degenerate: $g_{*}/u_{*}(u_{*}+1)=\varepsilon$.
However,
the anomalous dimension $\gamma_{f}^{*}=\varepsilon$ is independent of its
coordinate.

\begin{figure}
\begin{center}
\includegraphics[width=7cm]{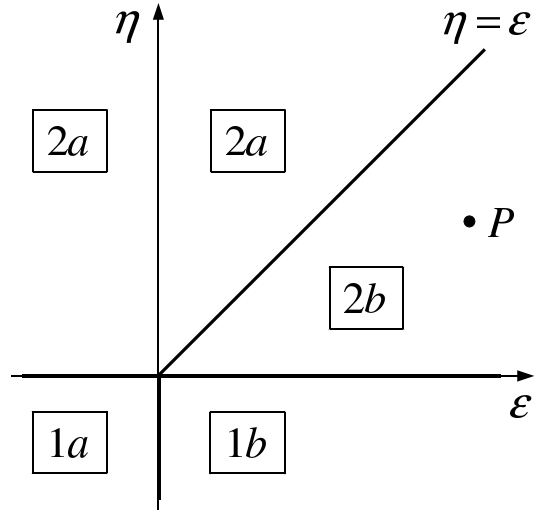}
\caption{\label{fig:patt}
Domains of IR stability of the fixed points in the model
(\protect\ref{velocF}). The numbers in boxes correspond to the fixed
points (1{\it a})--(2{\it b}) in the text. $P$ is the Kolmogorov point
$\varepsilon=2\eta=8/3$.}
\end{center}
\end{figure}

In figure~\ref{fig:patt} we show the domains in the $\varepsilon$--$\eta$
plane, where the fixed points listed above are IR attractive.
The boundaries of all domains are given by straight lines; there are
neither gaps nor overlaps between the domains. This fact is exact due
to the absence of the higher-order terms in the $\beta$ functions
(\ref{betaF}). We also stress that the Kolmogorov values of the exponents
$\varepsilon=8/3$, $\eta=4/3$ lie deep inside the domain of stability
of the nontrivial rapid-change point (2{\it b}); there is no borderline
going through this point.

The RG and OPE analysis of the preceding sections equally applies to the
model (\ref{velocF}) with a finite correlation time. The IR asymptotic
expressions for the structure functions in the nontrivial regime (2{\it b})
have the forms (\ref{OR})--(\ref{A2}) with dimensions (\ref{KDE}),
(\ref{2.6}) and (\ref{KDF}). For the regime (1{\it b}) one has to
replace $\xi$ with $\varepsilon$:
\begin{eqnarray}
\Delta_{\omega}=2, \quad \Delta_{\parallel} = 1+\varepsilon/2, \quad
\Delta_{\theta'}= d+1+\varepsilon/2, \quad \Delta_{m}=1,
\nonumber \\
\Delta [\theta^{n}] = n\Delta_{\theta} = -n, \quad
\Delta_{k,s} = s \varepsilon.
\label{DimF}
\end{eqnarray}
For the trivial regimes ({\it a}), the dimensions coincide with their
canonical values: $\Delta_{\omega}=2$, $\Delta_{\parallel} = 1$ and so on.
In these regimes, turbulent advection is irrelevant in the leading order
of the IR behaviour, and the difference between the frozen (1{\it a}) and
the rapid-change (2{\it a}) cases manifests itself only in correction terms
(for example, in UV corrections governed by the eigenvalues of the matrix
$\Omega$). Probably for this reason the difference between such regimes was
not mentioned, e.g., in \cite{AM} \cite{AM1}, \cite{Glimm}, \cite{Walls}.
Adopting the terminology of phase transitions, used in those papers,
we can say that a first-order transition occurs when the point in the
$\varepsilon$--$\eta$ plane, representing the state of the system, moves
continuously from domain (1{\it b}) to (2{\it b}) and crosses the line
$\eta=0$, $\varepsilon>0$: the effective correlation time jumps
discontinuously from infinity to zero. On the other hand, if the correlation
time changes continuously and passes all finite values from infinity to
zero, the point in the $\varepsilon$--$\eta$ plane `stacks' on the line
$\eta=0$, $\varepsilon>0$ (we recall that the value of $u_{*}$ is arbitrary
for the special fixed point with $\eta=0$). It remains to note that the
critical dimensions change continuously from (\ref{DimF}) to
the `rapid-change' values (\ref{KDE}), (\ref{2.6}) and (\ref{KDF}).

\section{Conclusion} \label{sec:Conc}

We studied inertial-range behaviour of a passive scalar in a random shear
flow, modelled by a $d$-dimensional generalization of the Gaussian ensemble
introduced in \cite{AM}. The case of vanishing correlation time
(\ref{vello})--(\ref{veloc2}) is discussed in detail, but all the results
are generalized to the case of finite correlation time (\ref{velocF}) with
the spectrum ${\cal E} \propto k_{\bot}^{1-\varepsilon}$ and the dispersion
law $\omega \sim k_{\bot}^{2-\eta}$.

It turns out that possible nontrivial types of the IR behaviour reduce
to the two limiting cases: the rapid-change type behaviour, realized
$\varepsilon>\eta>0$, and the frozen (time-independent) behaviour,
realized for $\varepsilon>0$, $\eta<0$.
The structure functions in the IR range exhibit scaling
behaviour of the form (\ref{ScS}), and the corresponding dimensions
are found exactly: (\ref{KDE}) for the rapid-change case and (\ref{DimF})
for the frozen case.

The resulting inertial-range asymptotic expressions, presenting
the main outcome of this study, are summarized in (\ref{ScC})--(\ref{A2}).

In a few respects, the IR behaviour of the model differs drastically
from that of the isotropic Kraichnan's rapid-change model:

(i) The scaling is strongly anisotropic in the sense that the critical
dimensions, related to the directions parallel and perpendicular to the
flow, are different.

(ii) The structure functions in the IR range $r \gg \ell$ retain the
dependence on the UV scale $\ell$ (the second Kolmogorov hypothesis
is violated).

(iii) Due to the absence of relevant dangerous operators with negative
dimensions in the corresponding OPE, the structure functions appear finite
for $r \ll {\cal L}$, where ${\cal L}$ is the integral scale,
and thus reveal no anomalous (multi)scaling in the sense of (\ref{HZ1}).
Thus the first Kolmogorov hypothesis is valid for our model.

(iv) For the finite-correlated isotropic case, the Kolmogorov values
$\varepsilon/2=\eta=4/3$ lie exactly on the crossover line between the
rapid-change and frozen regimes \cite{Chetak}--\cite{Juha2}. For the
present model, they lie inside the domain of the rapid-change regime;
there is no crossover line going through this point. This result is in
agreement with the analysis of \cite{Glimm} and in disagreement
with \cite{AM}--\cite{AM2}. Possible explanation of the discrepancy
between existing results was proposed in \cite{Walls}: it was argued that
existence of a crossover line going through the Kolmogorov point depends
on the specific choice of the model parameters: if the amplitudes in the
velocity correlators are related to the IR scale, the crossover disappears.
The possibility that the stability domains of fixed points indeed change
if the IR scale is introduced into the velocity correlators, was also
discussed within the RG approach to the isotropic model; see section
VIII in the e-print version of \cite{RG3}.

In this connection we stress that we found no crossover line, going through
the Kolmogorov point, although in our approach the amplitudes in the velocity
correlators (which play the part of the coupling constants) are related to
the UV scale, see (\ref{g0}) and (\ref{kopls}).
In the terminology of \cite{Walls}, the correlators are generalized at the
dissipation length. It is this choice that provides the agreement between
the RG and other approaches to Kraichnan's model (for the discussion
and comparison of various approaches, see \cite{AnttiPaolo}). In particular,
if the UV scale $\Lambda\sim 1/\ell$ was replaced by the IR scale
$m \sim 1/{\cal L}$ in (\ref{g0}), the invariant charge (\ref{InC}) in the
inertial range would tend to zero instead of the nontrivial fixed point
$w_{*}$ from (\ref{FP}), and the IR behaviour of the model would be trivial
at all. Thus the problem requires further investigation.

Two concluding remarks are in order. Although our results are exact, they
are derived by RG and OPE resummations of the original perturbation series,
and, in principle, their range of validity can be restricted by some
boundaries in the $\xi$--$\eta$ plane, whose existence and location
cannot be determined within the perturbative approach itself. The behaviour
of the model can notably change, for example, for $\eta>2$ (where the
dispersion law becomes abnormal) or $\xi>2$ (where the eddy diffusivity
becomes IR divergent). On the other hand, we mostly studied the structure
functions, determined by the statistics of relative motion of the particles
in the flow. The behaviour of individual particles is more subtle and
sensitive to the details of the velocity statistics \cite{AM8}.

In order to understand deeper the difference in the IR behaviour of the
passive scalar in a weakly anisotropic velocity ensemble \cite{Uni} and
the shear flow of the type \cite{AM}, it would be desirable to construct
a more general model, which included them both as special limiting cases.
Such a model is expected to demonstrate some kind of crossover between the
IR behaviour described above for the latter case, and the anomalous
(multi)scaling behaviour for the former one. It is not yet clear, however,
how to do this. Probably, the idea \cite{AM4} to approximate the isotropic
case by a family of shear flows averaged with respect to their shearing
directions will be useful here. This work remains for the future.

\section*{Acknowledgments}
The authors are indebted to L.Ts.~Adzhemyan, Michal Hnatich, Juha Honkonen,
Antti Kupiainen and Paolo Muratore~Ginanneschi for discussions.
AVM was supported in part by the Dynasty Foundation.
NVA thanks the Department of Theoretical Physics in the University of
Helsinki for their warm hospitality in Summer 2010.

\end{document}